# Observing the Sun with the Murchison Widefield Array

*D. Oberoi[1], R. Sharma[1], S. Bhatnagar[2], C. J. Lonsdale[3], L. D. Matthews[3], I. H. Cairns[4], S. J. Tingay[5], L. Benkevitch[3], A. Donea[6], S. M. White[7], G. Bernardi[8], J. D. Bowman[9], F. Briggs[10], R. J. Cappallo[3], B. E. Corey[3], A. Deshpande[11], D. Emrich[5], B. M. Gaensler[4,12], R. Goeke[13], L. J. Greenhill[14], B. J. Hazelton[15], M. Johnston-Hollitt[16], D. L. Kaplan[17], J. C. Kasper[18], E. Kratzenberg[3], M. J. Lynch[5], S. R. McWhirter[3], D. A. Mitchell[19,12], M. F. Morales[15], E. Morgan1[3], A. R. Offringa[10], S. M. Ord[5], T. Prabu[11], A. E. E. Rogers[3], A. Roshi[20], J. E. Salah[3], N. Udaya Shankar[11], K. S. Srivani[11], R. Subrahmanyan[11,12], M. Waterson[5], R. B. Wayth[5], R. L. Webster[21,12], A. R. Whitney[3], A. William[5], C. L. Williams[13]*

[1]National Centre for Radio Astrophysics, Tata Institute of Fundamental Research, Pune, India (div@ncra.tifr.res.in)

[2]National Radio Astronomy Observatory, Socorro, NM, USA

[3]MIT Haystack Observatory, Westford, MA, USA

[4]University of Sydney, Sydney, Australia

[5]Curtin University, Perth, Australia

[6]Monash University, Melbourne, Australia

[7]Air Force Research Laboratory, Kirtland, NM, USA

[8]Square Kilometre Array South Africa (SKA SA), Cape Town, South Africa

[9]Arizona State University, Tempe, AZ, USA

[10]The Australian National University, Canberra, Australia

[11]Raman Research Institute, Bangalore, India

[12]ARC Centre for Excellence for All-sky Astrophysics (CAASTRO)

[13]MIT Kavli Institute for Astrophysics and Space Research, Cambridge, MA, USA

[14]Harvard-Smithsonian Center for Astrophysics, Cambridge, MA, USA

[15]University of Washington, Seattle, WA, USA

[16]Victoria University of Wellington, New Zealand

[17]University of Wisconsin--Milwaukee, Milwaukee, WI, USA

[18]University of Michigan, Ann Arbor, MI, USA

[19]CSIRO Computational Informatics, Marsfield, Australia

[20]National Radio Astronomy Observatory, Charlottesville, WV, USA

[21]The University of Melbourne, Melbourne, Australia

## Abstract

The Sun has remained a difficult source to image for radio telescopes, especially at the low radio frequencies. Its morphologically complex emission features span a large range of angular scales, emission mechanisms involved and brightness temperatures. In addition, time and frequency synthesis, the key tool used by most radio interferometers to build up information about the source being imaged is not effective for solar imaging, because many of the features of interest are short lived and change dramatically over small fractional bandwidths.

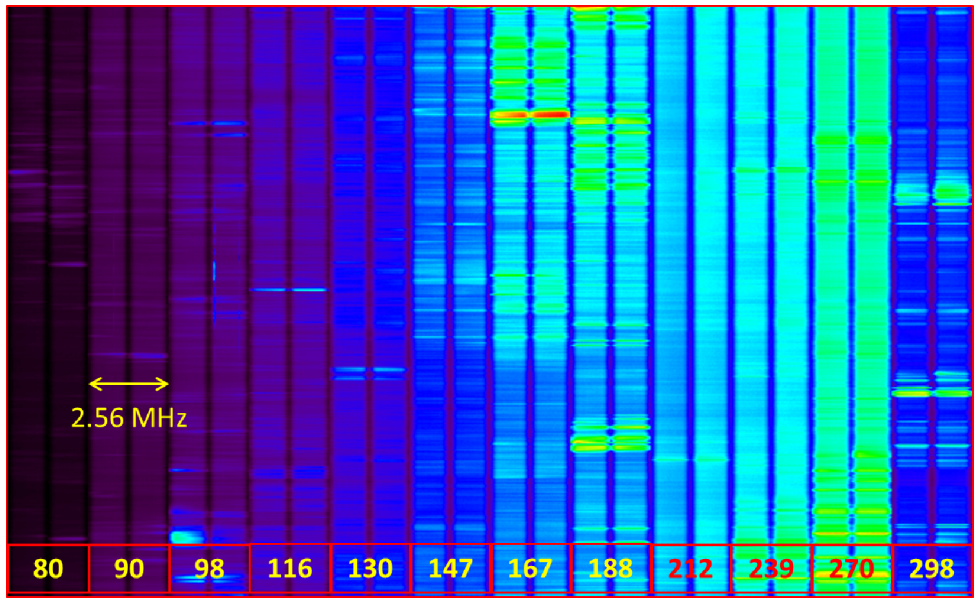

**Figure 1.** The figure shows an example raw dynamic spectrum from MWA solar observations. Time runs along the y axis and 300s of data are shown. The MWA observing band was divided into 12 roughly log-spaced groups of 2.56 MHz each spanning the entire 80-300 MHz observing band and are shown along the x axis. The numbers in the boxes give the central frequency of a given 2.56 MHz chunk. The colour shows the auto-correlation amplitude for the XX polarisation for one of the tiles. The time and spectral resolution are 0.5 s and 40 kHz respectively. These data have not been corrected for the instrumental bandpass. The bandpass is quite smooth across individual groups, but the uncorrected bandpass does contribute to the jumps across groups. The vertical dark stripes are instrumental in origin.

Building on the advances in radio frequency technology, digital signal processing and computing, the kind of instruments needed to simultaneously capture the evolution of solar emission in time, frequency, morphology and polarization over a large spectral span with the requisite imaging fidelity, and time and frequency resolution have only recently begun to appear. Of this class of instruments, the Murchison Widefield Array (MWA) is best suited for solar observations. The MWA has now entered a routine observing phase and here we present some early examples from MWA observations.

## 1. Introduction

The Murchison Widefield Array (MWA) is precursor for the Square Kilometre Array (SKA) and is located at the Australian site chosen for the SKA. It uses a novel design based on elements comprising 16 dual polarization dipoles. Referred to as tiles, 128 of these elements are spread over a region 3 km in diameter with a very strong central condensation. The MWA operates in the 80-300 MHz spectral range and the detailed engineering design is described in [1]. Its design is optimised for a few targeted science applications including solar imaging. The MWA science case is presented in [2] and further details of heliospheric science in [3]. The MWA commenced routine observing in the second half of 2013. It is now available for observing to the entire user community. Details about proposing and observing are available at http://mwatelescope.org.

## 2. Examples of early solar observations with the MWA

Since the 32 element MWA Prototype, the MWA solar images have represented the state-of-the-art in high fidelity and high dynamic range solar imaging. MWA data have also revealed the presence of weaker short lived and narrow band non-thermal emission than what is usually detected by most low frequency radio spectrographs [4]. Here we provide some examples of recent MWA observations to illustrate its ability to capture the details of solar emission.

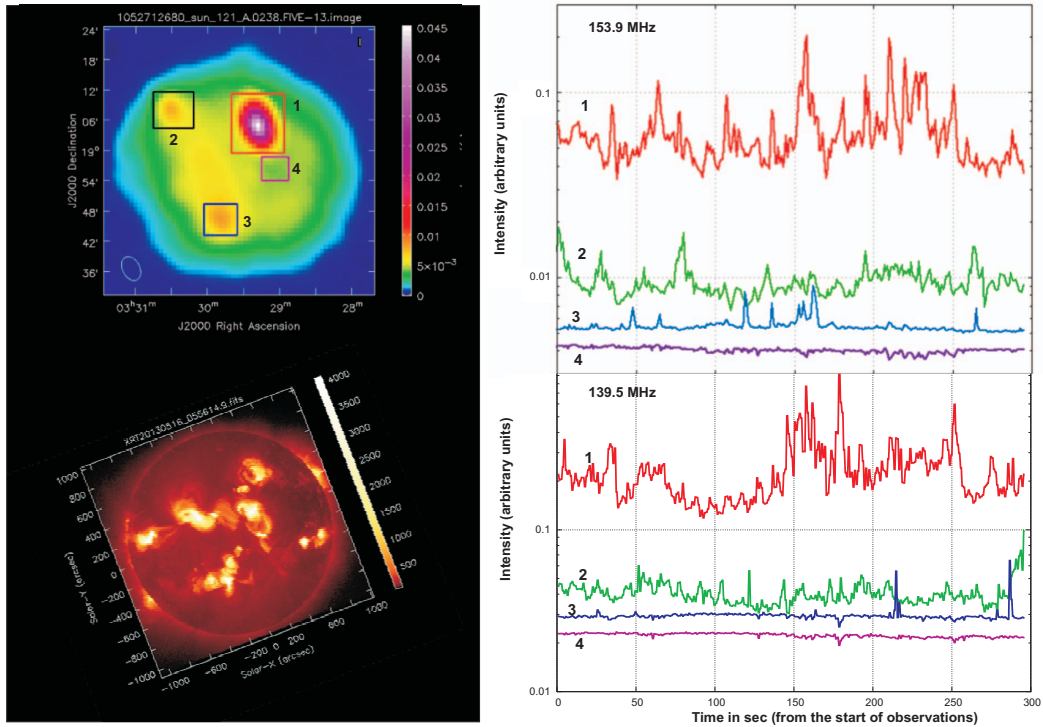

**Figure 2**. The top left panel shows a MWA radio image of the Sun at 153.9 MHz, corresponding to a time and frequency averaging of 1 s and 640 kHz, respectively, using data taken on May 16, 2013 04:15:02 UT. The dynamic range of this image is ~1400. The bottom left panel shows an X-ray image of the Sun close to the same time from the XRT instrument, onboard the Hinode spacecraft. The X-ray image has been rotated to align it with the orientation of the radio image. The panels on the right show the light curves corresponding to the features in the boxes marked 1 through 4 in the top left panel and 153.9 MHz (top) and 139.5 MHz (bottom). The light curves span ~300 s with a time resolution of 1 s. The labelled boxes have been chosen to pick sites of progressively weaker emission.

Figure 1 shows an example dynamic spectrum of the total power (auto-correlation) in the XX polarisation from one of the MWA elements. It demonstrates the ability of the MWA to simultaneously sample the entire 80-300 MHz band by organizing its 30.72 MHz of available bandwidth into smaller groups spread across the band. Presence of low level short lived emission features which change abruptly across groups of contiguous spectral bands are clearly seen. The level of activity is seen to be higher in some parts of the band as compared to the others. These data correspond to a period of low activity on the Sun and are weaker than the features usually seen in radio spectrographs.

An example of how this low level variability manifests itself in the image plane is illustrated in Fig 2. This figure shows the light curves of observed intensity variations in the brightest features present in the marked regions of the image for two different frequencies, 139.5 and 153.9 MHz. Though there is a region which shows much more variability than any other (~600%) there are a number of other regions which also show variability at varying levels. It is also evident that the light curves at two frequencies separated by 10% fractional bandwidth are qualitatively very different, demonstrating the narrow band nature of these emissions. The data used for Fig. 2 come from a period of low solar activity.

Figure 3 illustrates the changes in the nature of the emission between the quiescent and more active emission as seen in the uv-plane. For the quiescent Sun, the observed cross-correlations (visibilities) follow the expected profile for an extended source of size a little larger than the optical solar disc. In this particular instance, in the presence of active emission, the visibilities are completely dominated by this partially resolved non-thermal emission.

## 3. Acknowledgements

This scientific work makes use of the Murchison Radio-astronomy Observatory, operated by CSIRO. We acknowledge the Wajarri Yamatji people as the traditional owners of the Observatory site. Support for the MWA comes from the U.S. National Science Foundation (grants AST-0457585, PHY-0835713, CAREER-0847753, and AST-0908884), the Australian Research Council (LIEF grants LE0775621 and LE0882938), the U.S. Air Force Office of Scientific Research (grant FA9550-0510247), and the Centre for All-sky Astrophysics (an Australian Research Council Centre of Excellence funded by grant CE110001020). Support is also provided by the Smithsonian Astrophysical Observatory, the MIT

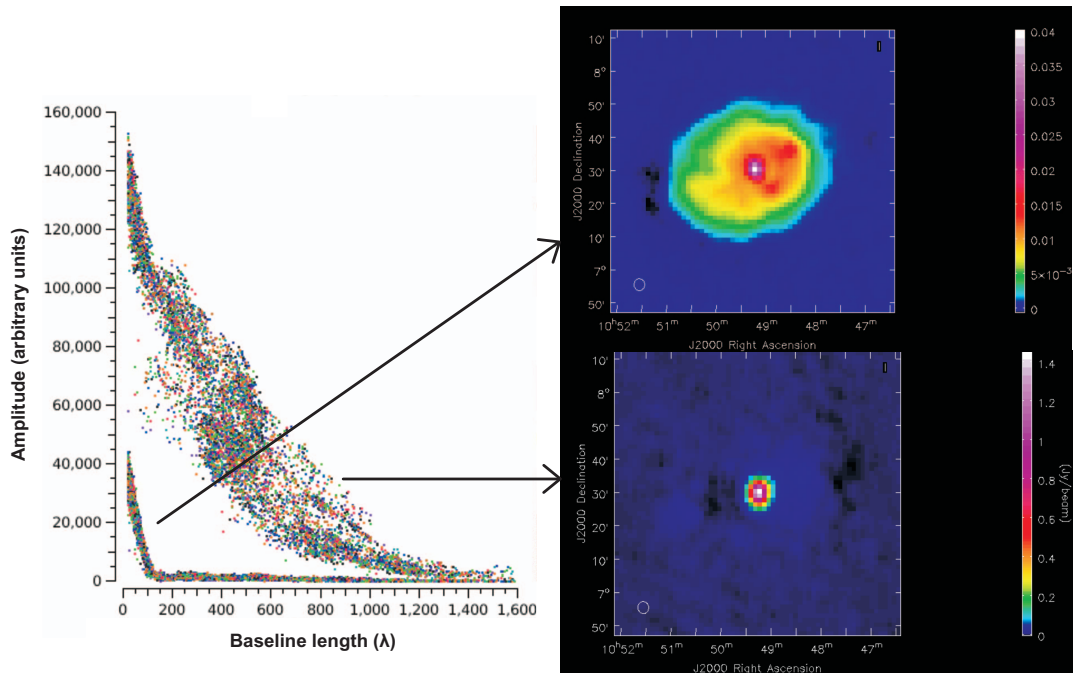

**Figure 3.** The left panel shows the amplitudes of the observed visibilities as a function of the length of the baseline in lambda for a single spectral slice 40 kHz wide and centered at 232.26 MHz for data taken on September 3, 2013. The panel shows data for two different 0.5 s time slices about a minute apart. The lower observed values correspond to the quiescent solar emission, and the signature of the extended disc of the Sun is clearly evident in the visibilities dropping rapidly with increasing baseline length. The higher observed values correspond to a 0.5 s time slice when the Sun was more active. This emission is dominated by a partially resolved source. The right panels show the images for the quiescent Sun (top) and the active Sun (bottom). The brightest point in the active Sun image is a factor of ~40 brighter than the corresponding point in the quiescent Sun image. The dynamic range of the active Sun image is insufficient to show the disc of quiescent thermal solar radio emission in presence of the very bright non-thermal emission.

School of Science, the Raman Research Institute, the Australian National University, and the Victoria University of Wellington (via grant MED-E1799 from the New Zealand Ministry of Economic Development and an IBM Shared University Research Grant). The Australian Federal government provides additional support via the Commonwealth Scientific and Industrial Research Organisation (CSIRO), National Collaborative Research Infrastructure Strategy, Education Investment Fund, and the Australia India Strategic Research Fund, and Astronomy Australia Limited, under contract to Curtin University. We acknowledge the iVEC Petabyte Data Store, the Initiative in Innovative Computing and the CUDA Center for Excellence sponsored by NVIDIA at Harvard University, and the International Centre for Radio Astronomy Research (ICRAR), a Joint Venture of Curtin University and The University of Western Australia, funded by the Western Australian State government.